\documentclass[aps,prd,twocolumn,showpacs,superscriptaddress,groupedaddress]{revtex4}

\usepackage{graphicx}  
\usepackage{dcolumn}   
\usepackage{bm}        
\usepackage{amssymb}   
\usepackage{xcolor}
\usepackage{hyperref}

\hypersetup{colorlinks=true, linkcolor=blue, citecolor=blue}

\begin{document}
\title{Galactic axions search with a superconducting resonant cavity}

\author{D. Alesini}
\affiliation{INFN, Laboratori Nazionali di Frascati, Frascati (Roma), Italy}
\author{C. Braggio}
\affiliation{INFN, Sezione di Padova, Padova, Italy}
\affiliation{Dip.\,di Fisica e Astronomia,  Padova, Italy}
\author{G. Carugno}
\affiliation{INFN, Sezione di Padova, Padova, Italy}
\affiliation{Dip.\,di Fisica e Astronomia,  Padova, Italy}
\author{N. Crescini}
\affiliation{INFN, Laboratori Nazionali di Legnaro,  Legnaro (PD), Italy}
\affiliation{Dip.\,di Fisica e Astronomia,  Padova, Italy}
\author{D. D'\,Agostino}
\affiliation{Dip.\,di Fisica E.R. Caianiello, Fisciano (SA), Italy and INFN, Sez. di Napoli, Napoli, Italy}
\author{D. Di Gioacchino}
\affiliation{INFN, Laboratori Nazionali di Frascati, Frascati (Roma), Italy}
\author{R. Di Vora}
\email[Corresponding author ]{divora@pd.infn.it}
\affiliation{INFN, Sezione di Padova, Padova, Italy}
\author{P. Falferi}
\affiliation{Istituto di Fotonica e Nanotecnologie, CNR Fondazione Bruno Kessler, I-38123
Povo, Trento, Italy}
\affiliation{INFN, TIFPA,  Povo (TN), Italy}
\author{S. Gallo}
\affiliation{INFN, Sezione di Padova, Padova, Italy}
\affiliation{Dip.\,di Fisica e Astronomia,  Padova, Italy}
\author{U. Gambardella}
\affiliation{Dip.\,di Fisica E.R. Caianiello, Fisciano (SA), Italy and INFN, Sez. di Napoli, Napoli, Italy}
\author{C. Gatti}
\email[Corresponding author ]{claudio.gatti@lnf.infn.it}
\affiliation{INFN, Laboratori Nazionali di Frascati, Frascati (Roma), Italy}
\author{G. Iannone}
\affiliation{Dip.\,di Fisica E.R. Caianiello, Fisciano (SA), Italy and INFN, Sez. di Napoli, Napoli, Italy}
\author{G. Lamanna}
\affiliation{Dip.\,di Fisica and INFN, Sez. di Pisa, Pisa, Italy}
\author{C. Ligi}
\affiliation{INFN, Laboratori Nazionali di Frascati, Frascati (Roma), Italy}
\author{A. Lombardi}
\affiliation{INFN, Laboratori Nazionali di Legnaro,  Legnaro (PD), Italy}
\author{R. Mezzena}
\affiliation{INFN, TIFPA,  Povo (TN), Italy}
\affiliation{Dip.\,di Fisica, Povo (TN), Italy}
\author{A. Ortolan}
\affiliation{INFN, Laboratori Nazionali di Legnaro,  Legnaro (PD), Italy}
\author{R. Pengo}
\affiliation{INFN, Laboratori Nazionali di Legnaro,  Legnaro (PD), Italy}
\author{N. Pompeo}
\affiliation{Dept. of Engineering, University Roma Tre, Rome, Italy}
\author{A. Rettaroli}
\email[Corresponding author ]{alessio.rettaroli@lnf.infn.it}
\affiliation{INFN, Laboratori Nazionali di Frascati, Frascati (Roma), Italy}
\affiliation{Dip.\,di Matematica e Fisica Universit\`a di Roma 3, Roma, Italy}
\author{G. Ruoso}
\affiliation{INFN, Laboratori Nazionali di Legnaro,  Legnaro (PD), Italy}
\author{E. Silva}
\affiliation{Dept. of Engineering, University Roma Tre, Rome, Italy}
\author{C. C. Speake}
\affiliation{School of Physics and Astronomy, Univ. of Birmingham, Birmingham, United Kingdom}
\author{L. Taffarello}
\affiliation{INFN, Sezione di Padova, Padova, Italy}
\author{S. Tocci}
\affiliation{INFN, Laboratori Nazionali di Frascati, Frascati (Roma), Italy}

\begin{abstract}
To account for the dark matter content in our Universe, post-inflationary scenarios predict  for the  QCD axion a mass in the range $(10-10^3)\,\mu\mbox{eV}$. Searches with haloscope
experiments in this mass range require the monitoring of
resonant cavity  modes with frequency above 5\,GHz, where several experimental limitations occur due to linear amplifiers, small volumes,  and low quality factors of Cu resonant cavities.
In this paper we deal with the last issue, presenting the result of a search for galactic axions using a haloscope based on a $36\,\mbox{cm}^3$ NbTi superconducting cavity.
The cavity worked at $T=4\,\mbox{K}$ in a 2\,T magnetic field and exhibited a quality factor $Q_0= 4.5\times10^5$ for the TM010 mode at 9\,GHz. With such values of $Q$ the axion signal is
significantly increased with respect to copper cavity haloscopes. Operating this setup we set the limit $g_{a\gamma\gamma}<1.03\times10^{-12}\,\mbox{GeV}^{-1}$ on the axion
photon coupling for a mass of about 37\,$\mu$eV.
A comprehensive study of the NbTi cavity at different magnetic fields, temperatures, and frequencies is also presented.
\end{abstract}
\maketitle

\section{Introduction}
\label{sec:introduction}
The axion is a pseudoscalar particle predicted by S.\,Weinberg\,\cite{Weinberg} and F.\,Wilczek\,\cite{Wilczek} as a consequence of the
mechanism introduced by R.D.\,Peccei and H.\,Quinn\,\cite{PecceiQuinn} to solve the ``strong CP problem''.
Axions are also well motivated dark-matter (DM) candidates with expected mass laying in a broad range from peV to few meV\,\cite{PDG2018}. Post-inflationary scenarios restrict this range to $(10-10^3)\,\mu\mbox{eV}$\,\cite{PDG2018}, where a rich experimental program will probe the axion existence in the next decade. Among the experiments, ADMX\,\cite{ADMX}, HAYSTAC\,\cite{HAYSTAC}, ORGAN\,\cite{ORGAN}, CULTASK\,\cite{CULTASK} and KLASH\,\cite{KLASH} will use a haloscope, i.e. a detector composed of a resonant cavity immersed in a strong magnetic field as proposed by P.\,Sikivie\,\cite{Sikivie}.
When the resonant frequency of the cavity $\nu_c$ is tuned to the corresponding axion mass $m_ac^2/h$, the expected power deposited by DM axions is given by\,\cite{HAYSTAC}
	\begin{equation}
	\label{eq:power}
	P_{a}=\left( g_{\gamma}^2\frac{\alpha^2}{\pi^2}\frac{\hbar^3 c^3\rho_a}{\Lambda^4} \right) \times
	\left( \frac{\beta}{1+\beta} \omega_c \frac{1}{\mu_0} B_0^2 V C_{mnl} Q_L \right),
	\end{equation}
where $\rho_a=0.45$\,GeV/cm$^3$ is the local DM density, $\alpha$ is the fine-structure constant, $\Lambda=78$\,MeV is a scale parameter related to hadronic physics, and $g_{\gamma}$ is a model dependent parameter equal to $-0.97$ $(0.36)$ in the KSVZ (DFSZ) axion model\,\cite{KSVZ,DFSZ}.
It is related to the coupling appearing in the Lagrangian $g_{a\gamma\gamma}=(g_{\gamma}\alpha/\pi\Lambda^2)m_a$. The second parentheses contain the vacuum permeability $\mu_0$, the magnetic field strenght $B_0$, the cavity volume $V$, its angular frequency $\omega_c=2\pi\nu_c$, the coupling between cavity and receiver $\beta$ and the loaded quality factor $Q_L=Q_0/(1+\beta)$, where $Q_0$ is the unloaded quality factor; here $C_{mnl}\simeq O(1)$ is a geometrical factor depending on the cavity mode.

The axion mass range studied by haloscopes up to now is limited to few $\mu$eV.
Exploring larger ranges at higher values requires the excitation of modes with frequency above a few GHz where several experimental limitations occur:
(\textit{i}) the technology of linear amplifier limits the sensitivity \,\cite{Lamoreaux}; (\textit{ii}) conversion volumes are smaller since the normal modes resonant frequencies are inversely proportional to the cavity radius; and (\textit{iii}) the anomalous skin effect reduces the copper cavities quality factor at high frequencies. Solutions to the first two issues are proposed for instance in\,\cite{Kuzmin} and\,\cite{JJeong}, respectively.

The optimum value of $Q$-factor for haloscopes is  $\sim10^6$, as estimated by the coherence time of DM axions\,\cite{QUAX1}.
A 10\,GHz copper cavity, cooled at cryogenic temperature, barely reaches $Q\sim10^5$, a value that rapidly decreases with increasing frequency. In this paper we present a substantial improvement obtained for the quality factor with a ``superconducting haloscope" composed of a superconducting cavity (SCC) operated in high magnetic fields.
This activity has been done within the QUAX experiment, which searches DM axions using a ferromagnetic haloscope\,\cite{QUAX1,QUAX2}. The same experimental apparatus can be used as a Sikivie's haloscope\,\cite{Sikivie} exploiting the TM010 mode of the cylindrical cavity. In this work we substitute the copper cavity with a superconducting one, to increase the quality factor and thus the measurement sensitivity. We refer to the Primakoff haloscope of the QUAX collaboration as ``QUAX$-a\gamma$''.

In Sec.\,\ref{sec:setup} we describe the characterization of the SCC and the measurement setup, while in Sec.\,\ref{sec:limits} we present the results of the axion search done by operating the SCC in magnetic field and the corresponding exclusion limit on the coupling $g_{a\gamma\gamma}$. Finally, in Sec.\,\ref{sec:conclusion} we discuss the future prospects of the QUAX$-a\gamma$ experiment for the Primakoff axion search.

\section{Experimental apparatus}
\label{sec:setup}

\subsection{Superconducting cavity}
\label{sec:cavity}
To increase the quality factor and match the optimal condition for the coupling to cosmological axions, it is natural to consider SCCs
as they were widely studied in accelerator physics. However, in axion searches these are operated in strong magnetic fields that, on
the one hand, weaken superconductivity and, on the other, are screened by the superconducting material. To overcome both these
limitations we designed a cavity divided in two halves, each composed by a Type II superconducting body and copper endcaps. Type II superconductors are infact known to have a reduced sensitivity to the applied magnetic field. Moreover, magnetic field penetration in the inner cavity volume may be facilitated by interrupting the screening supercurrents with the insertion of a thin ($30\,\mu$m) copper layer between the two halves.

	\begin{figure}
	  \begin{center}
	    \includegraphics[width=.45\textwidth]{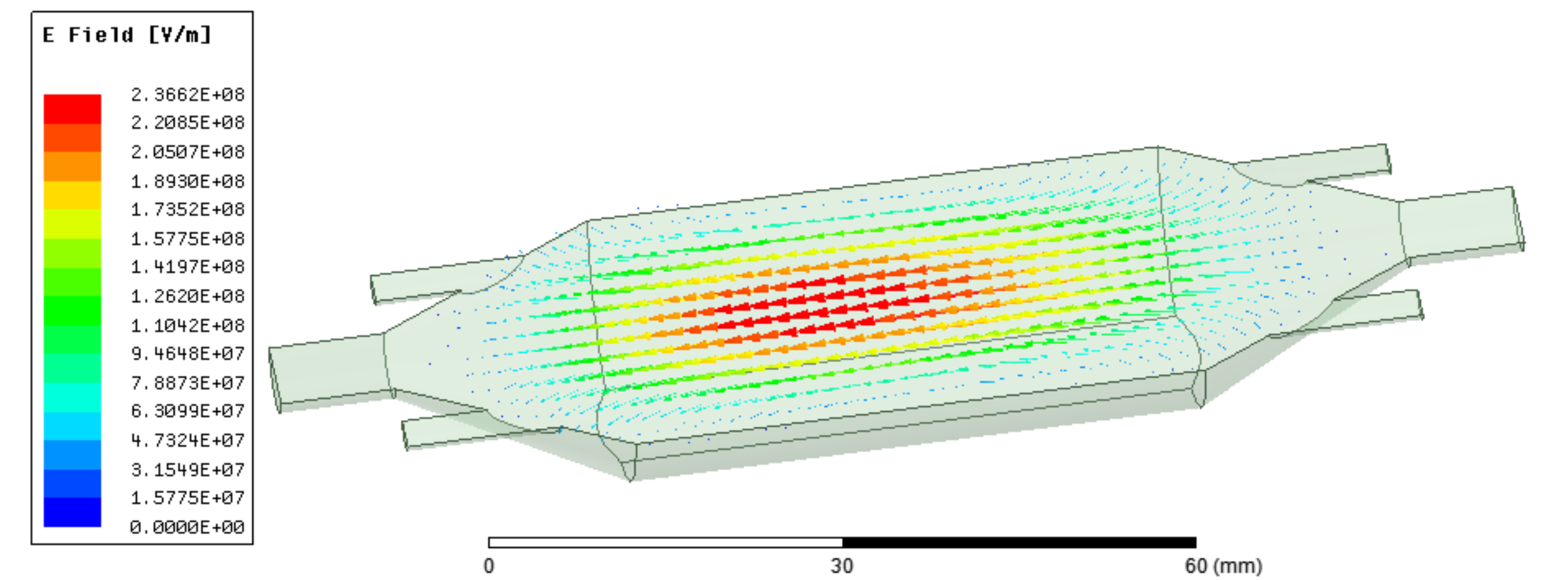}\\
	    \includegraphics[width=.4\textwidth]{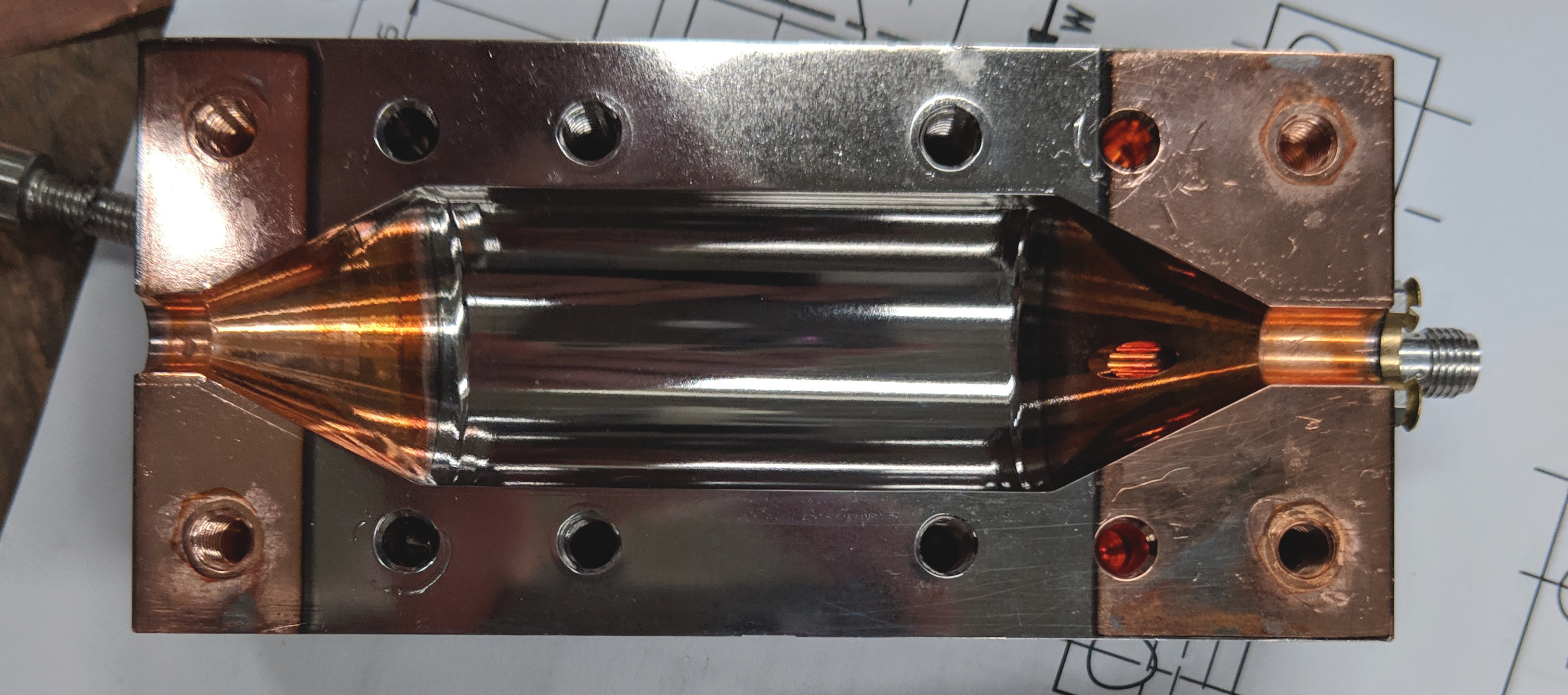}
	    \caption{The upper image represents the electric field of 9.08\,GHz TM010 mode in arbitary amplitude units, while the lower photo is one of the two halves of the superconducting cavity.}
	    \label{fig:quaxcavity}
	  \end{center}
	\end{figure}

The cavity layout is shown in the upper part of Fig.\,\ref{fig:quaxcavity}, featuring two identical copper semi-cells with cylindrical body and conical endcaps to reduce current dissipation at interfaces. The cylindrical body is 50\,mm long with diameter 26.1\,mm, while the cones are 19.5\,mm long. For the detection of axions through its interaction with the electron spin\,\cite{QUAX2}, part of the cylinder was flattened to break the angular symmetry and prevent the degeneration of mode TM110.
A finite element calculation performed with ANSYS HFSS\,\cite{HFSS} of the TM010 mode used for Primakoff axion detection gives a frequency $\nu_c^{\mathrm{sim}}=9.08$\,GHz and a coefficient $C_{mnl}=0.589$ in a volume $V=36.43~\mbox{cm}^3$. The calculated field profile of this mode is shown in Fig.\,\ref{fig:quaxcavity}.
Because of this hybrid geometry, the quality factor is expressed as
\begin{equation}
	\label{eq:qf}
	\frac{1}{Q_{0}}=\frac{R_s^\mathrm{cyl}}{G_\mathrm{cyl}}+\frac{R_s^\mathrm{cones}}{G_\mathrm{cones}},
\end{equation}
where $R_s$ are the surface resistances.
The simulation yields $G_\mathrm{cones}=6270.11\,\Omega$ and $G_\mathrm{cyl}=482.10\,\Omega$. At 9\,GHz and 4\,K temperature the surface resistance for Cu is $R_s^\mathrm{Cu}=4.9\,\mbox{m}\Omega$\,\cite{skineffect}.
A pure Cu cavity with this geometry would have $Q_0^\mathrm{Cu}\simeq9\times10^4$ while an hybryd cavity with copper cones and no losses on the cylindrical surface would have $Q_0^\mathrm{max}=G_\mathrm{cones}/R_s^\mathrm{Cu}\simeq1.3\times10^6$.

To test this promising simulation results, a prototype of the cavity was fabricated as shown in the lower part of Fig.\,\ref{fig:quaxcavity}.
The inner cylindrical part of the cavity was coated by means of a 10\,cm planar magnetron equipped with a NbTi target. The estimated coating thickness ranges between 3 to 4\,$\mu$m. To minimize the losses due to the interaction of fluxons\,\cite{abrikosov} with the superconducting microwave-current, only the cylindrical body, where the currents of the mode TM010 are parallel to the applied field, were coated as evidenced by the different colors of the lower picture of Fig.\,\ref{fig:quaxcavity}.

\begin{figure}
	  \begin{center}
	    \includegraphics[width=.5\textwidth]{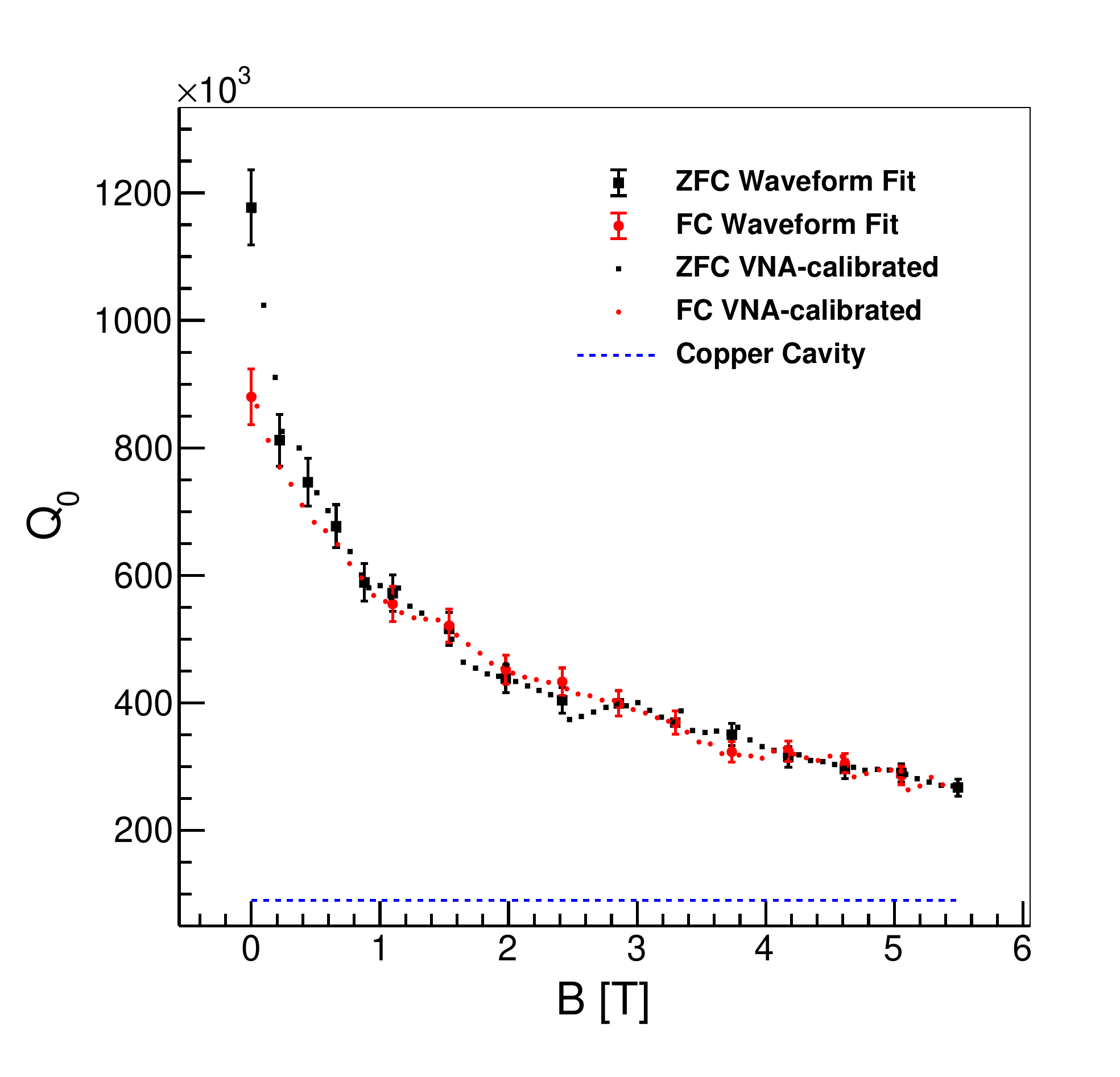}
	    \caption{Unloaded quality factor $Q_0$ vs external magnetic field amplitude $B$ for the hybrid Cu-NbTi cavity, compared with the one of a Cu cavity (blue horizontal line). ZFC points are measured by cooling the cavity before increasing the field, the opposite procedure was used for FC data (see text for details). Unloaded quality factors $Q_0$ indicated by large markers with error bars were derived from fits to the reflection and transmission waveforms, $S_{11}(\nu)$ and $S_{12}(\nu)$. These measurements were used to correct the loaded quality factors $Q_L$ measured by the VNA, shown, after the correction, as small markers without error bars.}
	    \label{fig:q0vsb}
	  \end{center}
	\end{figure}

We characterized the cavity in a thermally controlled gas-flow cryostat equipped with an 8\,T superconducting magnet located at Laboratori Nazionali di Frascati (LNF). No copper layer was inserted between the two halves. Two tunable antennas were coupled to the cavity mode and connected through coax cables to a Vector Network Analyzer for the measurement of the reflection and transmission waveforms, $S_{11}(\nu)$ and $S_{12}(\nu)$. The unloaded quality factor $Q_0$ was extracted from a simultaneous fit of the two waveforms. An expected systematic error of $\pm5\%$ follows from the fit procedure.
At 4.2\,K and no applied external magnetic field we measured $Q_0=1.2\times10^6$ in agreement with the
maximal expected value $Q_0^\mathrm{max}$ and corresponding to a surface resistance of the NbTi $R_s^\mathrm{NbTi}=(20\pm20)\,\mu\Omega$.
We repeated the measurement for different values of the applied magnetic field in the range $0-5$\,T both in zero-field cooling (ZFC)
and field cooling (FC), thus reducing the temperature of the cavity without or with external magnetic field, respectively. The results are shown in Fig.\,\ref{fig:q0vsb}.
For $B=2$\,T, the nominal field used in our axion search, we measured $Q_0^{2\,\mathrm{T}}=4.5\times10^5$, a factor $\sim5$ better than a bulk Cu cavity; at 5\,T we measured $Q_0^{5\,\mathrm{T}}=2.95\times10^5$, a factor $\sim3.3$ better than a Cu cavity.
Comparing FC and ZFC measurements, we observe a systematic difference below about 0.5\,T due to magnetic field trapping in the superconducting material. For higher values the measurements agree within the errors showing that the magnetic field is able to penetrate the cavity walls and that the superconductor is in the flux flow state\,\cite{BardeenStephen}.

\begin{figure}[htbp]
	    \includegraphics[width=.5\textwidth]{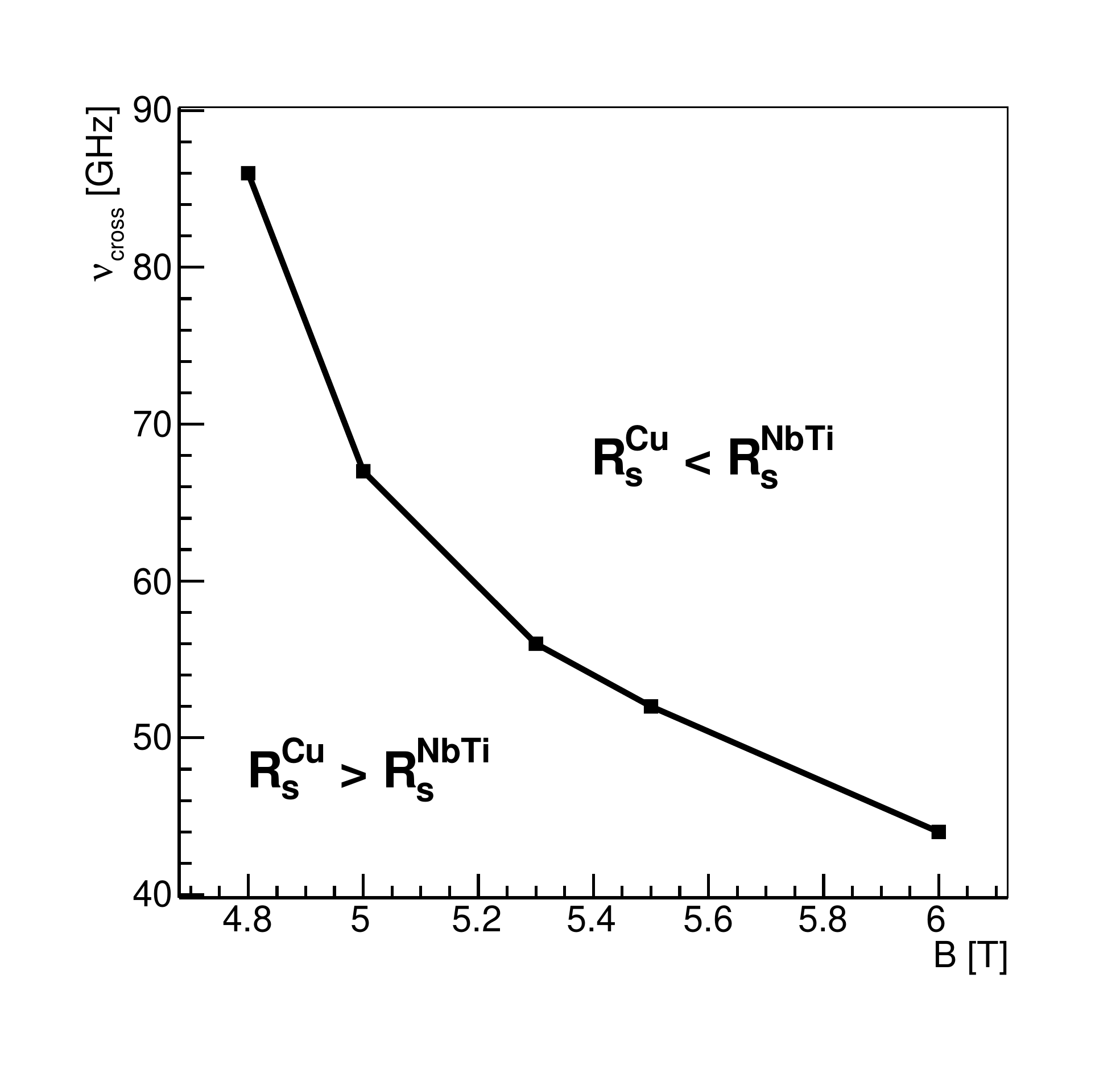}
	    \caption{Estimated crossing frequency as a function of applied $B$ field.}
	    \label{fig:crossing}
	\end{figure}

In a recent analysis\,\cite{ASC2018}, the variation of the surface resistance of this cavity with the magnetic field was interpreted
taking into account the vortex-motion contribution within the Gittleman and Rosenblum (GR) model\,\cite{GittlemanR,PompeoSilva}.
In particular, the depinning frequency\,\cite{GittlemanR} was estimated to be about 44\,GHz.
Below this frequency, losses due to vortex motion are suppressed, while they are maximal at higher frequency where the flux-flow resistivity
reaches the value $\rho_\mathrm{ff}=c_\mathrm{ff}\rho_n B/B_{c2}$. Here, $B$ is the applied DC field, $c_\mathrm{ff}=0.044$ is a correction taking into account
mutual orientation of fluxons and microwave currents, $\rho_n=7.0\times10^{-7}\,\Omega\mbox{m}$ is NbTi resistivity in the normal state and $B_{c2}=13\,\mbox{T}\times\left(1-(T/T_c)^2\right)$ is the temperature dependent critical field\,\cite{ASC2018}.
Comparing the surface resistance derived from the GR model with these parameters and the anomalous surface resistance of Cu we estimated, for different values of the applied field, the crossing frequency for which the losses in NbTi are equal to the losses in Cu. The result is shown in Fig.\,\ref{fig:crossing}. At $T=4.2$\,K the NbTi cavity is expected to show lower losses up to an applied field of about 4.5\,T. For higher fields the crossing frequency rapidly decreases down to 45\,GHz for $B=6$\,T.
By lowering the temperature down to 100 mK we expect a 20\% improvement of the $Q_0$ and larger values for the crossing frequency. In fact, at this temperature the critical field $B_{c2}$ reaches its maximum value, 13\,T, reducing the flux-flow resistivity.

\subsection{Magnet and readout electronics}
\label{sec:readout}
\begin{figure}[htbp]
	  \begin{center}
	    \includegraphics[width=.25\textwidth]{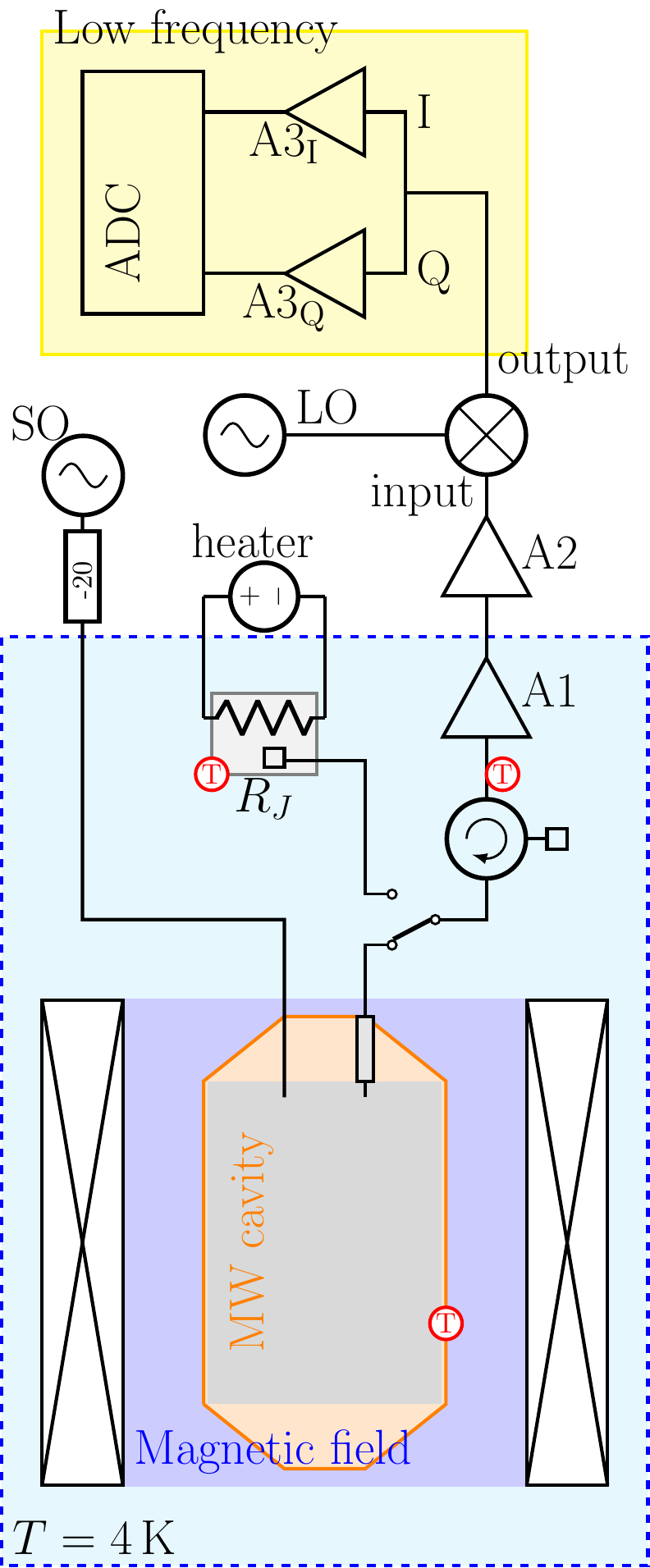}
	    \caption{QUAX$-a\gamma$ setup: the blue dashed line encloses the liquid helium temperature part of the apparatus, the yellow rectangle stands for the low-frequency electronics and the red circled $T$s represent the thermometers. See text for further details.}
	    \label{fig:quaxsetup}
	  \end{center}
	\end{figure}

A replica of the cavity described in Sec.\,\ref{sec:cavity} was mounted in the experimental site at Laboratori Nazionali di Legnaro (LNL), which hosts an apparatus capable to search for galactic axions\,\cite{QUAX2}. The scheme of the apparatus is shown in Fig.\,\ref{fig:quaxsetup}. The SCC is inside a vacuum chamber inserted in a superconducting magnet. The magnet is a NbTi compensated solenoid, 15\,cm bore and 50\,cm height, generating a central field of 2\,T with homogeneity better than 20\,ppm on a 20\,mm-long line along the central axis. A superconducting switch is installed to perform measurements also in persistent current mode. The bias current of 50\,A is supplied by a high-stability current generator. The magnet and the vacuum chamber are immersed in a LHe bath at the temperature of 4.2\,K.
The cavity is instrumented with two antennas, a weakly coupled one was used to inject probe signals in the cavity with a source oscillator (SO). The second antenna, with a variable coupling, is connected through a coax cable to a low noise cryogenic HEMT amplifier (A1) with gain $G_1\simeq40$\,dB. Before reaching the amplifier the coax cable is connected to a cryogenic switch and then to an isolator. The switch, used for calibration purposes, allows the replacement of the cavity output with the output of a resistor ($R_J$).
The temperature of the resistor is kept constant by a heater and read by a thermometer.
The setup is completed by a second FET amplifier (A2) at room temperature with gain $G_2\simeq39$\,dB and a down-converter mixer referenced to a local oscillator (LO).
The in-phase (I) and quadrature (Q) components of the mixer output are further magnified by two identical room temperature amplifiers (A3$_{I,Q}$) with $G_3\simeq50$\,dB each and acquired by a 16 bit ADC sampling at 2\,MHz. The acquisition program controls both the oscillators, the ADC and the applied magnetic field.
Three thermometers monitor the temperature of the cavity, of the resistance $R_J$ and of the amplifier A1, with typical temperatures of $T_c=4.3\,$K, $T_J=4.5\,$K and $T_a=5.1\,$K, respectively.

\section{Experimental results}
\label{sec:limits}
With the LNL setup described in Sec.\,\ref{sec:setup} we performed the first search for galactic axions using a SCC.
The frequency of the TM010 cavity mode, $\nu_c=9.07$\,GHz, was in good agreement with the simulated value $\nu_c^{\mathrm{sim}}=9.08$\,GHz.
A copper layer was used to interrupt circular screening-supercurrents and allow magnetic field penetration. No impact was observed on the quality factor, since the longitudinal microwave currents of the TM010 mode are unaffected by the interposed mask. The penetration of the magnetic field in the cavity volume was verified by means of a 1\,mm YIG sphere positioned on the cavity axis with a teflon holder. Hybridization of the sphere ferromagnetic-resonance with the TM110 mode at 14\,GHz occurs with a field of 0.5\,T, showing a typical double resonance curve whose Lorentzian peaks have linewidths equal to the average of the YIG and cavity linewidths. The observation of this effect confirmed that the field distortotion were at the level of 100\,ppm or lower over the sphere volume. Measurements in the dispersive regime, i.\,e. with ferromagnetic resonance frequency different from the cavity one, also gave similar results.

 To down-convert and acquire the signal, the frequency of the local oscillator fed to the mixer was set to $\nu_\mathrm{LO}=\nu_c-500\,\mbox{kHz}$, and the $I$ and $Q$ components were combined to extract the right part of the down-converted spectra, where the cavity resonance lies.
We determined the total gain $G$ and the noise temperature $T_n$ of the amplification chain by heating the resistor $R_J$ from about 4.5 to 8.5\,K and by measuring the temperature and the corresponding Johnson noise\,\cite{Pino}. The resulting values are
	\begin{eqnarray}
	 	\label{eq:calibration}
	  T_n &=& \left(11.0\pm0.1\right)\,\mbox{K},
	  \\\nonumber
	  G &=& \left(1.96\pm0.01\right)\times10^{12}.
	\end{eqnarray}

\begin{figure}[htbp]
	  \begin{center}
	    \includegraphics[width=.5\textwidth]{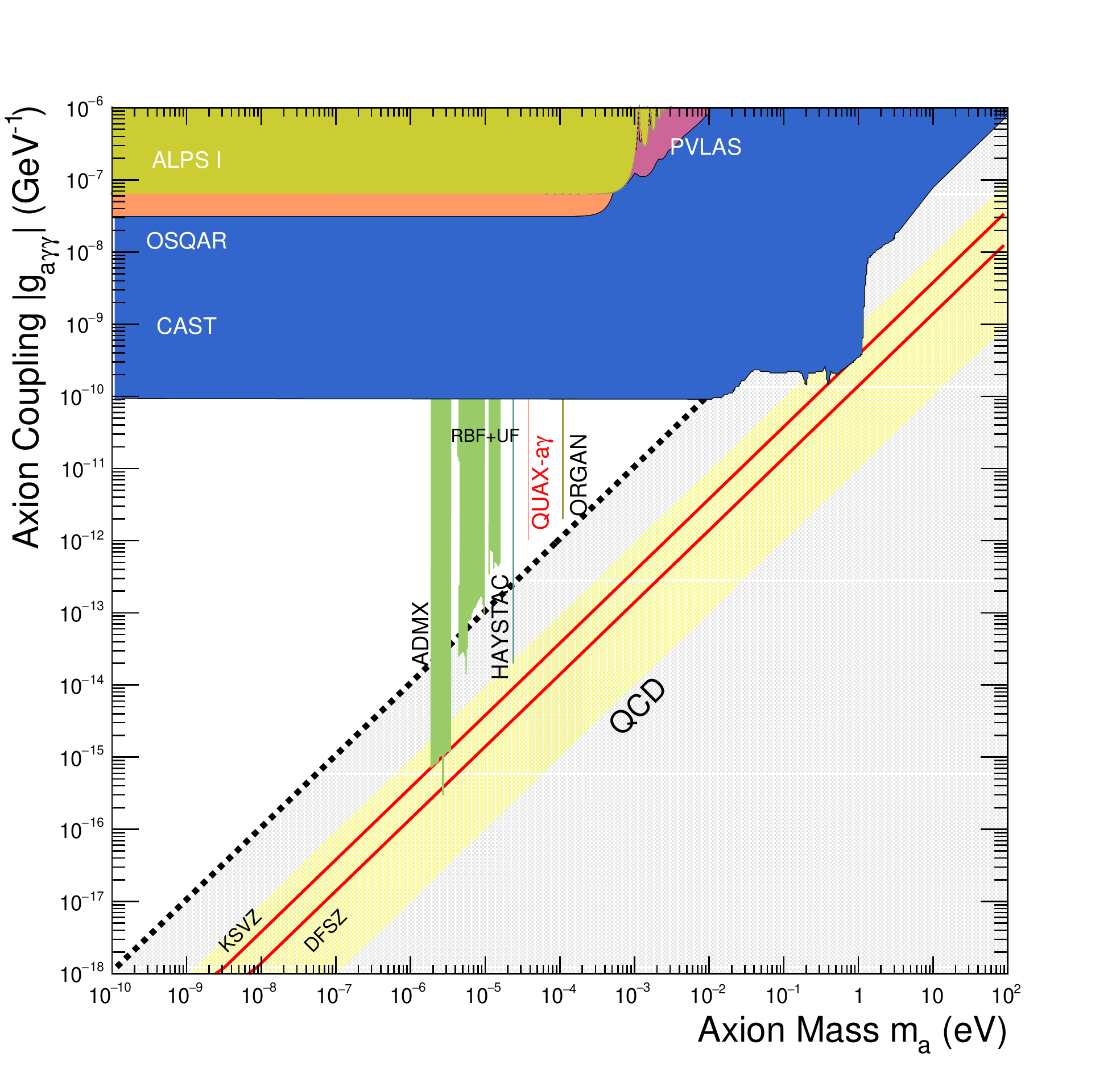}
	    \caption{Exclusion plot of the axion-photon coupling. The red lines with yellow error-band show the theoretical predictions for the KSVZ and DFSZ axions\,\cite{KSVZ,DFSZ}. The grey area shows the prediction form other hadronic axions models\,\cite{Nardi}. The experimental limits are mesured with light shining through a wall experiments\,\cite{ALPS, OSQAR}, from changes in laser polarization\,\cite{PVLAS} helioscopes\,\cite{CAST} and haloscopes\,\cite{ADMX,HAYSTAC,ORGAN,UF,BNL}, as the one in the present work.}
	    \label{fig:limits}
	  \end{center}
	\end{figure}

We set the magnetic field to 2\,T and measured the cavity quality factor $Q_0=4.02\times10^5$, compatible within the errors with our previous measurment. Finally, we critically coupled the tunable antenna and measured the loaded quality factor $Q_L=2.01\times10^5$.
With high quality factors the temperature stability of the system is a critical issue: in fact we observed a drift of the cavity resonance frequency of the order of the linewidth in the timescale of an hour. Thus the integration time was limited to $\Delta t=20$\,min.
	\begin{figure}[htbp]
	  \begin{center}
	    \includegraphics[width=.5\textwidth]{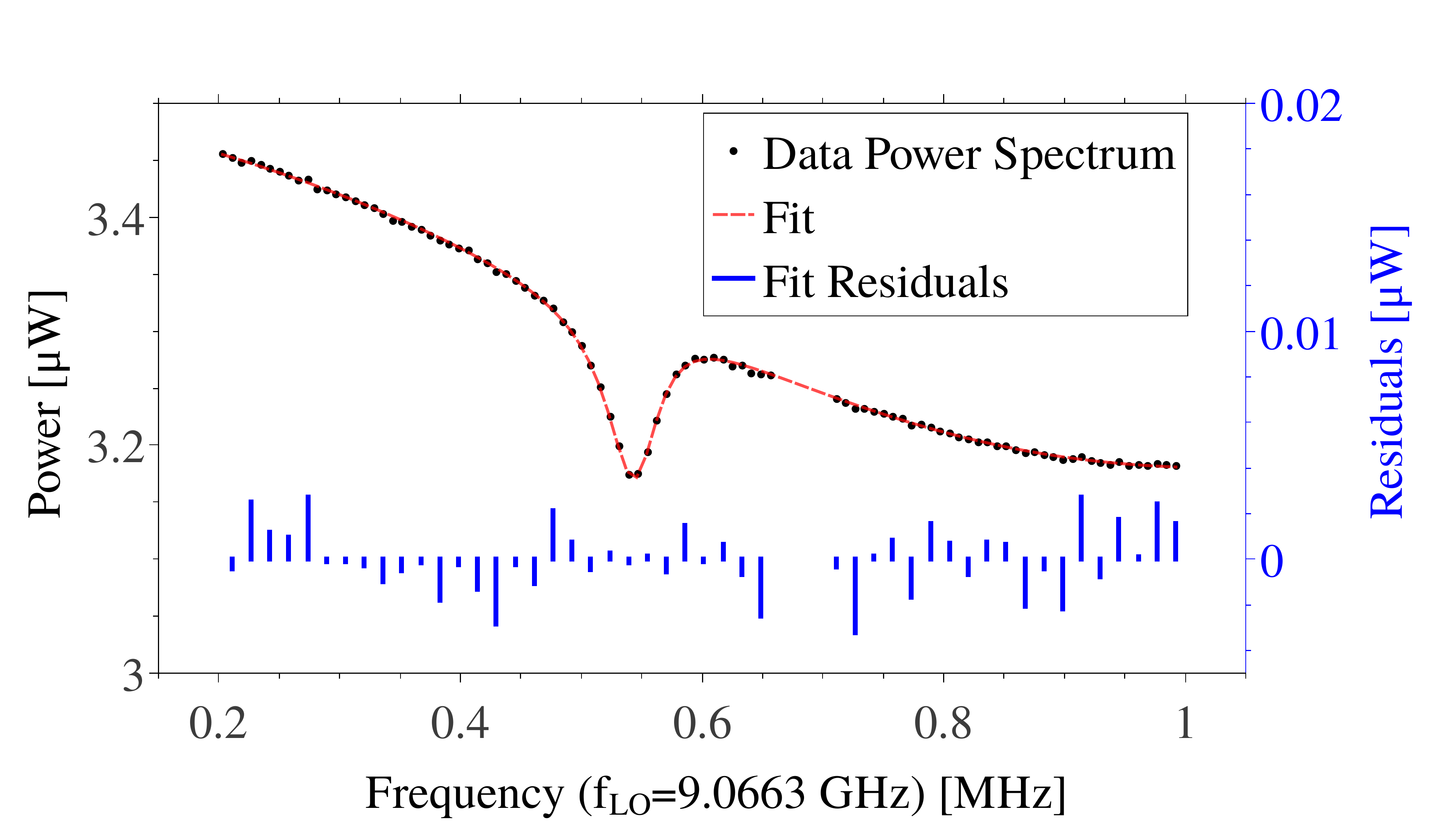}
	    \caption{Down-converted rf power at the ADC input. The collected data are the black dots (the errors are within the symbol dimension), the red-dashed line is the fit with the residuals reported in blue. A part of the bandwidth was removed due to systematic disturbances.}
	    \label{fig:power}
	  \end{center}
	\end{figure}
The collected data were analyzed with a FFT (Fast Fourier Transform) with a resolution bandwidth of $\Delta\nu=7812.5$\,Hz close to the axion linewidth, to maximize the SNR in a single bin.
The resulting 9375000 FFTs were RMS averaged, and the bins with disturbances introduced by the low-frequency electronics were excluded from the analysis.
The power spectrum was fit using a degree 5 polynomial to account for the off-resonance part of the spectra, which are due to the non-uniform gain of ADC and amplifiers.
We modeled the on resonance spectrum with the absorption profile of a Lorentzian curve. Since the temperature of the cavity was lower than the one of the isolator, the spectrum of power reflected by the cavity shows a minimum at the resonance frequency.
The measured spectrum with fit and residuals are shown in Fig.\,\ref{fig:power}.
The residuals are distributed according to a Gaussian probability density function and their standard deviation scales as $\sqrt{\Delta t}$ as expected.
To get the equivalent power at the cavity output we divide the power measured at the ADC input by the total measured gain $G$ in Eq.\,\ref{eq:calibration}.
Its standard deviation is $\sigma_P=6.19\,\times10^{-22}$\,W. This value is compatible with the prediction of the Dicke relation\,\cite{Dicke}
	\begin{equation}
	      \sigma_P=k_B T_S \sqrt{\frac{\Delta\nu}{\Delta t}}\simeq 5.5\times10^{-22}\,\mbox{W},
	\end{equation}
where $T_S=T_n+T_c=15.3$\,K.
The expected power generated by KSVZ axions in our cavity, derived from Eq.~\ref{eq:power}, is
	\begin{eqnarray}
	  \nonumber
	  &&P_\mathrm{a}=1.85\times10^{-25}\,\mbox{W} \left( \frac{V}{0.036\,\mbox{l}}\right) \left( \frac{B}{2\,\mbox{T}}\right)^2\left( \frac{g_{\gamma}}{-0.97}\right)^2
	  \\\nonumber
	  && \left( \frac{C}{0.589}\right)\left( \frac{\rho_a}{0.45\,\mbox{GeV}\mbox{cm}^{-3}}\right) \left( \frac{\nu_c}{9.067\,\mbox{GHz}}\right) \left( \frac{Q_L}{201000}\right).
	\end{eqnarray}
The 95\% single sided confidence limit (1.64$\sigma$), shown in Fig.\,\ref{fig:limits}, is $g_{a\gamma\gamma}<1.03\times10^{-12}\,\mbox{GeV}^{-1}$ in a frequency band of 45\,kHz at $\nu_c$ corresponding to a mass range of $\sim0.2$\,neV
around $m_{a}\simeq37.5\,\mu$eV.

\section{Conclusions}
\label{sec:conclusion}
SCCs appear as a natural choice for high frequency haloscopes as their quality factor matches the one of cosmological axions. In this work, we characterized a SCC by testing it under a high magnetic field at cryogenic temperature. After successful tests, we performed a single-mass axion search, extracting an upper limit on $g_{a\gamma\gamma}$ for a narrow frequency band. This result is limited by the low magnetic field, the high system temperature and the small cavity volume. A new experimental setup is now in preparation consisting of a dilution refrigerator, a quantum limited Josephson Parametric Amplifier (JPA) and an 8~T superconducting magnet. At 50~mK with a quantum limited amplifier, such as a JPA, the system temperature is reduced to about 400~mK. Operating a 20~cm long NbTi-cavity in a 5~T magnetic field, we expect, from our measurements, a quality factor $Q_0^{5\,\mathrm{T},50\,\mathrm{mK}}=2.95\times1.2\times10^5=3.54\times10^5$, a factor 4 better than a copper cavity. With this setup, the expected 95\% exclusion limit would be $g_{a\gamma\gamma}<4\times10^{-14}\,\mbox{GeV}^{-1}$ for $m_{a}\simeq37.5\,\mu$eV a value that touches the region expected for KSVZ axions.

\section{Acknowledgements}
We would like to thank M. Iannarelli and G. Pileggi for their support in the preparation of the LNF setup and F. Tabacchioni and M. Martini for their help with the cryogenic system. We also thank E. Berto, F. Calaon, M. Tessaro, M. Zago and M. Rebeschini for their work on the mechanics and electronics of the LNL setup, L. Castellani and G. Galet for building the magnet current source and N. Toniolo, M. Gulmini and S. Marchini for the help with the DAQ system. Finally, we would like to thank A. Benato for mechanical construction of the resonant cavities and C. Pira for the deposition of the NbTi film.


\begin{thebibliography}{99}
\bibitem{Weinberg} S.\,Weinberg, Phys. Rev. Lett. {\bf 40}, 223 (1978).
\bibitem{Wilczek} F.\,Wilczek, Phys. Rev. Lett. {\bf 40}, 279 (1978).
\bibitem{PecceiQuinn} R.\,D.\,Peccei and H.\,R.\,Quinn, Phys. Rev. Lett. {\bf 38}, 1440 (1977); Phys. Rev. D {\bf 16}, 1791 (1977).
\bibitem{PDG2018} A.\,Ringwald, L.J.\,Rosenberg and G.\,Rybka, ``Axions and other similar particles,'' in Particle Data Group Phys. Rev. D 98, 030001 (2018).
\bibitem{ADMX} S.\,Asztalos et al., Phys. Rev. D {\bf 64}, 092003 (2001). S.J. Asztalos et al., Phys. Rev. Lett. 104, 041301 (2010).
\bibitem{HAYSTAC} S.\,Al Kenany et al., Nucl. Instrum. Methods A {\bf 854}, 11 (2017); B.M.\,Brubaker et al., Phys. Rev. Lett. {\bf 118}, 061302 (2017).
\bibitem{ORGAN} B.T.\,McAllister et al., arXiv:1706.00209 [physics.ins-det].
\bibitem{CULTASK} W.\,Chung, PoS CORFU 2015, 047 (2016).
\bibitem{KLASH} D.\,Alesini et al., arXiv:1707.06010 [physics.ins-det]. C.\,Gatti et al., Contributed to the 14th Patras Workshop on Axions, WIMPs and WISPs, DESY in Hamburg, June 18 to 22, 2018, arXiv:1811.06754 [physics.ins-det].
\bibitem{Sikivie} P.\,Sikivie, Phys. Rev. Lett. {\bf 51}, 1415 (1983); Phys. Rev. D {\bf 32}, 2988 (1985).
\bibitem{KSVZ} J.\,Kim, Phys. Rev. Lett. {\bf 43}, 103 (1979); M.A.\,Shifman, A.I.\,Vainshtein, and V.I.\,Zakharov, Nucl. Phys. B {\bf 166}, 493 (1980).
\bibitem{DFSZ} M.\,Dine, W.\,Fischler, and M.\,Srednicki, Phys. Lett. B {\bf 104}, 199 (1981); A.P.\,Zhitnitskii, Sov. J. Nucl. Phys. {\bf 31}, 260 (1980).
\bibitem{Lamoreaux} S.K.\,Lamoreaux et al., ``Analysis of single-photon and linear amplifier detectors for
microwave cavity dark matter axion searches,'' Phys. Rev. D 88, 035020 (2013).
\bibitem{Kuzmin} L.\,Kuzmin et al., ``Single Photon Counter based on a Josephson Junction at 14 GHz for searching Galactic Axions,'' IEEE Trans. App. Sup. DOI: 10.1109/TASC.2018.2850019.
\bibitem{JJeong} J.\,Jeong et al., ``Concept of multiple-cell cavity for axion dark matter search,'' Phys.Lett. B {\bf 777}, 412 (2018).
\bibitem{QUAX1} R.\,Barbieri et al., ``Searching for galactic axions through magnetized media: The QUAX proposal,'' Phys. Dark Univ. {\bf 15}, 135 (2017).
\bibitem{QUAX2} N.\,Crescini et al.,``Operation of a ferromagnetic axion haloscope at m$_a$ = 58$\mu$eV,'' Eur. Phys. J. C (2018) 78:703.
\bibitem{HFSS} https://www.ansys.com/products/electronics/ansys-hfss
\bibitem{skineffect} G.E.H.\,Reuter and E.H.\,Sondheimer, Proc. R. Soc. A 195 (1948) 336. 
\bibitem{abrikosov} A.A.\,Abrikosov, ``The magnetic properties of superconducting alloys". Journal of Physics and Chemistry of Solids. 2 (3) (1957) 199–208
\bibitem{BardeenStephen} J.\,Bardeen and M.J.\,Stephen,``Theory of Motionof Vortices in Superconductors ,'' Phys. Rev. {\bf 140} (1965) A1197.
\bibitem{ASC2018} D.\,Di Gioacchino et al.,``Microwave losses in a dc magnetic field in superconducting cavities for axion studies,'' IEEE Trans. Appl. Sup. {\bf 29}, no. 5, (2019).
\bibitem{GittlemanR} J.\,I.\,Gittleman and B.\,Rosenblum,”Radio-Frequency Resistance in the Mixed State for Subcritical Currents” Physical Review Letters, vol.16, no.17, pp.734-736, Apr. 1966.
\bibitem{PompeoSilva} N.\,Pompeo and E.\,Silva, “Reliable determination of vortex parameters from measurements of the microwave complex resistivity,” Phys. Rev. B, vol. 78, no. 9, p. 094503, Sep. 2008.
\bibitem{Pino} C.\,Braggio et al.,``Characterization of a low noise microwave receiver for the detection of vacuum photons,'' Nucl. Inst. Meth. {\bf A603} (2009) 451-455.
\bibitem{Fano} A.E.\,Miroshnichenko et al.,``Fano resonances in nanoscale structures,'' Rev. Mod. Phys. {\bf 82}, 2257.
\bibitem{Dicke} R.H.\,Dicke, Rev. Sci. Instrum. {\bf 17} (7), 268 (1946).
\bibitem{Nardi} L.\,Di Luzio, F.\,Mescia, and E.\,Nardi, ``Redefining the Axion Window,'' Phys. Rev. Lett. {\bf 118}, 031801 (2017).
\bibitem{ALPS} K.\,Ehret et al. (ALPS Collab.), Phys. Lett. B {\bf 689}, 149 (2010).
\bibitem{OSQAR} R.\,Ballou et al. (OSQAR Collab.), Phys. Rev. D{\bf 2}, 092002 (2015).
\bibitem{PVLAS} F.\,Della Valle et al. (PVLAS Collab.), Eur. Phys. J. C{\bf 76}, 24 (2016).
\bibitem{CAST} M.\,Arik et al. (CAST Collab.), Phys. Rev. D{\bf 92}, 021101 (2015).
\bibitem{BNL} S.\,DePanfilis et al., Phys. Rev. Lett. {\bf59} (1987) 839; W.\,Wuensch et al., Phys. Rev. D{\bf40} (1989) 3153.
\bibitem{UF} C.\,Hagmann et al., Phys. Rev. D{\bf 42} (1990) 1297–1300.

\end{thebibliography}
\end{document}